\documentclass[sn-mathphys-num]{sn-jnl}

\usepackage{graphicx}%
\usepackage{multirow}%
\usepackage{amsmath,amssymb,amsfonts}%
\usepackage{amsthm}%
\usepackage{mathrsfs}%
\usepackage[title]{appendix}%
\usepackage{xcolor}%
\usepackage{textcomp}%
\usepackage{manyfoot}%
\usepackage{booktabs}%
\usepackage{algorithm}%
\usepackage{algorithmicx}%
\usepackage{algpseudocode}%
\usepackage{listings}%

\usepackage{mathtools}
\usepackage{physics}

\begin{document}

\title[Article Title]{Quantum Properties of Bipartite Separable Mixed State for Ancilla-Assisted Process Tomography}


\author[1]{\fnm{Zhuoran} \sur{Bao}}\email{zhuoran.bao@mail.utoronto.ca}

\author[1]{\fnm{Daniel F. V.} \sur{James}}\email{dfvj@physics.utoronto.ca}

\equalcont{These authors contributed equally to this work.}

\equalcont{These authors contributed equally to this work.}

\affil[1]{\orgdiv{Dept. of Physics}, \orgname{University of Toronto}, \orgaddress{\street{255 Huron St}, \city{Toronto}, \postcode{M5S 1A7}, \state{ON}, \country{Canada}}}


\abstract{It has been shown that the entanglement between the system and ancillary states is not a strict requirement for performing ancilla-assisted process tomography(AAPT). Instead, from a theoretical point of view, it only requires that the system-ancilla state be faithful, which, in the qubit case, is the invertibility of a certain matrix representing the state. Our paper takes on the operational definition of faithfulness, i.e., a state is faithful if one can extract complete information about the quantum process, and we restrict the process to single-qubit operations on a two-qubit system-ancilla state. We present a theoretical analysis that connects the invertibility problem to the concept of Sinisterness, which quantifies the correlation between two qubits and can be generalized to bipartite systems formed by qubits for a certain class of states. Using Sinisterness, we derive a way of constructing two-qubit states that are guaranteed to be faithful and estimate the bound on the average error of the process featured by the condition number. Our analysis agrees that the maximally entangled states provided the smallest error amplification. Nevertheless, it maps out a numerical region where the advantage of the entanglement starts.}




\maketitle

\section{Introduction}\label{sec1}

Characterizing the effect of some unknown device is crucial for quantum information science and quantum metrology. For example, one needs to keep track of decoherence when performing a computation in the context of quantum information processing. Alternatively, in polarimetry, the interaction of the polarized light with a sample could indicate the nature of that sample. Such quantum devices can be completely characterized by \emph{quantum process tomography} \cite{Chuang, Poyatos, Pryde}. 
    
Over the past twenty years, various approaches have been developed to enhance the efficiency of quantum process tomography, focusing on directly extracting the process from post-processing measurements with the aid of controlled two-qubit gates. Such a protocol is known as the direct characterization of quantum dynamics (DCQD) \cite{Mohseni2006, Mohseni2007, Mohseni2008}. However, any method utilizing two-qubit controlled gates is highly problematic for linear optical systems. Hence, the only available technique is indirect quantum process tomography, including the standard quantum process tomography (SQPT) \cite{Chuang, Poyatos, Pryde, ShuklaMhesh} and ancilla-assisted process tomography (AAPT) \cite{Altepeter}. For an n-qubit system, SQPT requires 2\(^{2n}\) different linearly independent initial states as input, but AAPT requires only one type of initial state. Therefore, AAPT has an advantage in dealing with optical systems, as polarization-correlated states can be used to reduce the number and settings of waveplates and polarizers required for state initialization. A drawback is that one needs to correlate an ancilla with a dimension at least as large as the original state to extract the process \cite {Ariano}. Despite this drawback, AAPT can still be of interest, for it is postulated that AAPT allows one to extract the error that occurred in a quantum algorithm if the error is localized to a part of the system. Besides the practical aspect, AAPT is a paradigm for studying general non-classical correlations. The success of AAPT depends on the existence of non-classical two-qubit correlations, which are not limited by the scope of entanglement and should include some of the non-classical correlations that originate from the operators' non-communicative nature, which is often referred to as quantum discord \cite{Modi}. For example, many bipartite separable mixed states exhibit this type of non-classical correlation. Yet, no entanglement, and as we will demonstrate later in the paper, they are indeed candidates for performing AAPT.
    
The initial bipartite system used in AAPT is usually considered a maximally entangled state. However, when aiming to produce a pure entangled state, environmental noise causes decoherence. The fidelity of a multi-qubit entangled state, depending on the size of the system, is much lower than perfect \cite{TP, ZL, VS}. In practice, we often deal with mixed entangled and even separable states, rather than pure states. So, the behaviour and effectiveness of these states performing AAPT are of theoretical interest. It has been shown that entanglement is not strictly required for AAPT; instead, one needs the state to be faithful \cite{Mohseni2008, Ariano, Altepeter, SHL, LZD, Wood}. According to D'Ariano and Presti's definition, a state is faithful if every different quantum channel maps the state to a different output state, and faithfulness is equivalent to the invertibility of the state's corresponding Jamiołkowski map \cite{Ariano, choi, jam}. A similar proof is also presented by Wood, Biamonte, and Cory \cite{Wood}. However, the proofs they provided were later shown to be incomplete by Lie and Jeong, who also completed the proof and proposed the concept of sensitivity, which features whether a change in the state can be sensed \cite{SHL}. Throughout the progress, the discussion on faithfulness largely remains abstract and mathematical, and few authors have mentioned a quantitative measure of faithfulness except D'Ariano and Presti, who noted that the measure should be proportional to the singular values of the Jamiołkowski map and used purity of the state as an example \cite{Ariano}. The reason is that any measure would depend on the method used to reconstruct the quantum channel. In optics, no such measure has been proposed yet. So, we would like to explore a measure associated with channel reconstruction based on the Pauli basis. We use the Pauli basis because it is the most widely used basis in polarization measurements, and results for other orthogonal qubit bases can be derived from the Pauli basis by rotations.

Here, we consider the application of Sinisterness to the problem. Sinisterness \cite{James} is a quantity characterizing two-qubit correlations using directly measurable quantities rather than a density operator reconstructed from tomography. By utilizing Sinisterness, one can separate the problem of whether AAPT will succeed from the problem of the exact amount of error amplification. It provides the best-case scenario estimate of the error amplification due to matrix inversion, and faithful two-qubit states are expected to have a non-zero Sinisterness.
    
The paper is organized into five sections. Section 2 briefly reviews the indirect process tomography, including the standard quantum process tomography (SQPT) and the ancilla-assisted process tomography (AAPT). Section 3 introduces the concept of Sinisterness and connects it to error amplification with various states. Section 4 provides the representation of arbitrary two-qubit separable mixed states and applies the concept of Sinisterness to various states, including maximally entangled states, general separable mixed states, and the Werner states. Finally, Section 5 is the conclusion, and a discussion on mixed states with more than one qubit in the system is provided in Appendix A.

\section{Background: Indirect Quantum Process Tomography} \label{sec:2}
In this section, we present both SQPT and AAPT as matrix multiplication equations for qubit systems and identify the invertibility problem relevant to the success of both strategies. The SQTP procedure has been discussed in detail in the work of Chuang and Nielson \cite{Chuang}. We briefly recap this strategy, allowing us to compare it to AAPT. Consider a d-dimensional Hilbert space, \(\mathcal{H}\), with a corresponding \(d^2\) dimensional operator space \(\mathcal{B(H)}\). An unknown process, which is not necessarily unitary, is applied to the system. The most general way for performing process tomography procedures is to appeal to the idea that a completely positive trace-preserving linear map can describe any quantum process applied to the system or part of the system:
\begin{equation}
\label{positivemap}
    \mathcal{E}(\hat{\rho}) = \sum_i {\hat{A_i}\hat{\rho}\hat{A}_i^{\dag}},
\end{equation}
where \(A_i\)'s satisfy \(\sum_i{ A_i^{\dag} A_i}=1\), and can be written as a linear combination of orthogonal Hermitian basis operators \(\hat{E}_i\) of \(\mathcal{B(H)}\). By orthogonality, we mean that the \(\hat{E_i}\)'s satisfy \(\Tr{\hat{E_i}^{\dag}\hat{E_j}}= d\delta_{ij}\). 

We choose to work with a system consisting of a single qubit, which is the most relevant to our analysis in the following sections. We also choose the basis to be the set of Pauli operators \(\hat{\sigma}_i\) since many properties of the Pauli operators are well known. Any other choice of orthogonal basis can be related to the Pauli operator by \(\hat{E}_i = \sum_{j} \epsilon_{ij}\hat{\sigma}_j\). The process is describe by a basis dependent superoperator, \( \chi\) \cite{Chuang} :
\begin{equation}
\label{epsilon(rho)}
    \mathcal{E} ( \hat{\rho} ) = \sum_{j,k=0}^{3} {\chi_{jk} \hat{\sigma_j} \hat{\rho} \hat{\sigma_k}^{\dag}}.
\end{equation}
To obtain the elements of the super operator \(\chi\) in SQPT, we need four linearly independent input states. Let these input states be \({\hat{\rho}^{(l)}}=\sum_{i=0}^{3}a_{i}^{(l)}\hat{\sigma}_i\).  Define the quantity \(B_{jikm}=\Tr{\hat{\sigma}_j\hat{\sigma}_i\hat{\sigma}_k\hat{\sigma}_m}\). We can write Eq. (\ref{epsilon(rho)}) as:
\begin{equation}
\label{epsilon(rho)3}
    \mathcal{E} ( \hat{\rho}^{(l)} )  = \sum_{i,j,k,m=0}^{3} { \chi_{jk}B_{jikm}a_{i}^{(l)}\hat{\sigma}_m}.
\end{equation}
But we can also expand the final state directly in the Pauli basis,
\begin{equation}\label{f}
    \mathcal{E} ( \hat{\rho}^{(l)} ) = \sum_{m=0}^3 a'^{(l)}_m \hat{\sigma}_m.
\end{equation}
Equating Eq. (\ref{f}) with Eq. (\ref{epsilon(rho)3}) :
\begin{equation}
    \label{epsilon(rho)4}
    a'^{(l)}_m = \sum_{i,j,k=0}^3\chi_{jk} B_{jikm} a_{i}^{(l)} = \sum_{i=0}^3\tilde{\chi}_{mi}a_i^{(l)} \longleftrightarrow a_{out} = \tilde{\chi}a_{in}.
\end{equation}
The equation above is a matrix multiplication equation if we view \(a'^{(l)}_m\), \( a^{(l)}_{i}\), and \(\Tilde{\chi}_{mi}\) as elements of four by four matrices \(a_{out}\), \(a_{in}\), and \(\tilde{\chi}\). The process encoded by \(\Tilde{\chi}\) can be found by making ensemble measurements to determine \(a_{out}\) and compute \(\tilde{\chi} = a_{out}a_{in}^{-1}\). Note that \(a_{in}\) is guaranteed to be invertible since it has linearly independent columns (the four linearly independent initial states). In general, for an n-qubit system, one must prepare \(2^{2n}\) different linearly independent initial states and then make as many as \(16^n\) projection measurements to figure out the final state \cite{Mohseni2008}. Additionally, more elaborate post-data processing is needed to extract the elements of \(\chi\) from \(\tilde{\chi}\) \cite{Pryde}. The above problems render SQPT quite ineffective in practice. 

By using correlations between an ancilla and the system, AAPT simplifies the initial state preparation step, as only one type of initial state is required. In short, observing Eq. (\ref{epsilon(rho)4}), AAPT searches for invertible four-by-four matrices that represent valid quantum states to replace \(a_{in}\). Since \(\tilde{\chi}\) is four by four, it implies that an ancilla qubit is needed. More generally, it has been demonstrated that the ancilla must have an associated Hilbert space at least as large as the system \cite{Mohseni2008, Altepeter}. The most often used system-ancilla states are the maximally entangled states. However, some separable mixed and non-maximally entangled states are also available for the procedure at the expense of increasing error amplification associated with matrix inversion \cite{Mohseni2006, Mohseni2007, Mohseni2008}.

D'Ariano and Preti proposed the concept of faithfulness \cite{Ariano}, which expresses that a system-ancilla state would be suitable for the AAPT procedure when every different process maps this state to a distinct final state. They proposed that faithfulness is equivalent to the existence of a one-to-one Jamiolkowski map. However, they proved only the sufficient condition of this equivalence. Later, Lie and Jeong completed the proof by proving the necessity \cite{SHL}. From an operational point of view, Mohseni, Rezakhani, and Lidar derived an equivalent condition of faithfulness in qubit systems and concluded that this condition for faithfulness is equivalent to the invertibility of a certain matrix representing the state in practice, functionally this matrix replaces \(a_{in}\) in Eq. (\ref{epsilon(rho)4}). A thorough derivation and description of the procedure can be found in reference \cite{Mohseni2008}. In our paper, we adopt Mohseni et al.'s notion, further explore this idea, and provide a measure for determining the effectiveness of performing AAPT with various separable mixed states. 

We start by defining the orthogonal basis for the system-ancilla operator space as \(\hat{\sigma}_i^{(A)} \otimes \hat{\sigma}_j^{(B)}\). Let \(\hat{\rho}_{AB}\) be the initial density matrix of the system-ancilla state. We can write \(\hat{\rho}_{AB}\) as:
\begin{equation}
\label{initial state}
    \hat{\rho}_{AB} = \sum_{i,j} \tau_{ij} \hat{\sigma_i}^{(A)}\otimes\hat{\sigma_j}^{(B)}
\end{equation}After the system undergoes the quantum process, the system-ancilla state is:
\begin{equation}
\label{final state}
\hat{\rho}'_{AB}=(\mathcal{E} \otimes I) (\hat{\rho}_{AB}) \equiv\sum_{ij} \tau_{ij}\sum_{lm}\chi_{lm}\hat{\sigma}_l^{(a)}\hat{\sigma}_i^{(a)}\hat{\sigma}_m^{(a)} \otimes \hat{\sigma}_j^{(b)}.
\end{equation}
Now, define \(B_{limk} = \Tr{\hat{\sigma}_l^{(a)}\hat{\sigma}_i^{(a)}\hat{\sigma}_m^{(a)}\hat{\sigma}_k^{(a)}}\) as before. We can rewrite Eq. (\ref{final state}) as below:
\begin{equation}
\begin{split}
\hat{\rho}'_{AB} & =\sum_{ijlm}\tau_{ij}\chi_{lm}(\sum_{k}B_{limk}\hat{\sigma}_k^{(a)})\otimes \hat{\sigma}_j^{(b)}\\
& =\sum_{kj} (\sum_{ilm}\tau_{ij}\chi_{lm}B_{limk})\hat{\sigma}_k^{(a)}\otimes \hat{\sigma}_j^{(b)}.
\end{split}
\end{equation}
But, we know that we can always write \(\hat{\rho}' =\sum_{kj} \tau'_{kj}\hat{\sigma}_k^{(a)}\otimes \hat{\sigma}_j^{(b)}\). Therefore, comparing the coefficients, we get:
\begin{equation}
\label{tau to tau'}
    \tau'_{kj} = \sum_{i,l,m}\chi_{lm} B_{limk}\tau_{ij}.
\end{equation}
We notice that Eq. (\ref{tau to tau'}) is identical to Eq. (\ref{epsilon(rho)4}) except we use different variables and identify \(\tau\) representing a quantum state. Then, to determine the process \(\tilde{\chi}=\sum_{l,m}\chi_{lm} B_{limk}\), the only condition required is that the matrix with elements \(\tau_{ij}\) to be invertible. The extraction of \(\chi\) from \(\tilde{\chi}\) can be done as in the SQPT case. The analysis above uses the Pauli basis and two qubits as an example. However, the scheme can be easily generalized to larger bipartite systems by replacing the Pauli basis with an orthogonal basis that spans the operator space. The condition for AAPT to succeed remains the same: one still needs the matrix \(\tau_{ij}\) to be invertible. Some examples of such mixed states are Werner and isotropic states \cite{Altepeter}. However, many other separable mixed states also work for AAPT. Therefore, we aim to develop a scheme that captures the full spectrum of faithful separable mixed states and suggest a recipe for constructing such states. Nevertheless, the experiment that demonstrates the superiority of maximally entangled states in terms of reduced error amplification uses only the separable Werner states as a test subject, representing all non-entangled states \cite{Mohseni2008}. We aim to find evidence that supports the theory.

\section{Sinisterness and Error Amplification}

In the previous section, we established that the faithfulness of a state is the same as the invertibility of its matrix representation \(\tau\), regardless of the size of the bipartite system. Thus, a straightforward condition for a state to be faithful is that \(\det(\tau) \neq 0\). The properties of \(\det(\tau)\) for a two-qubit system have been explored in detail in ref. \cite{James}; in particular, the sign of \(\det(\tau)\) reveals the handedness of the correlation tensor \(\langle (\hat{\sigma}_i - a_iI)\otimes (\hat{\sigma}_j-b_jI)\rangle \), where \(a_i, b_j\) comes from the coefficient in the decomposition:
\begin{equation}
    \hat{\rho}_{in} = \frac{1}{2}(I+\Vec{a}\cdot\hat{\Vec{\sigma}
    })\otimes\frac{1}{2}(I+\Vec{b}\cdot\hat{\Vec{\sigma}}) + \sum_{ij} c_{ij} \hat{\sigma_i}\otimes\hat{\sigma_j}.
\end{equation}
The left-handed parity states always have \(\det(\tau)<0\) and are entangled. Therefore, the quantity \(\det(\tau)\) is referred to as the Sinisterness of the state. Physically, Sinisterness quantitatively describes the correlations within the two-qubit states. For a pure state, det(\(\tau\))=-C\(^4\), where C is the concurrence. For a general mixed state, the Sinisterness is bound by functions of concurrence, which has been shown numerically \cite{James}. For separable mixed states, the absolute value of Sinisterness is bound by \(\frac{1}{27}\). For classically correlated states defined in the scope of quantum discord \cite{Modi}, the Sinisterness is zero, which AAPT will fail. Thus, we always need some level of quantumness to perform AAPT successfully. In the case of pure two-qubit states, it manifests as entanglement, but the situation becomes more complicated when mixtures are involved. An example of this is the mixed product state, which has a positive non-zero determinant, and the quantum nature comes from the non-commuting property of the Pauli operators.

Being invertible only guarantees the ability to find \(\tilde{\chi}\) in theory; in practice, when dealing with matrix inversion, one also requires the matrix to be well-conditioned \cite{Belsley}, so that the error amplification due to matrix inversion does not blow up. Recall that we defined \(\hat{\rho}_{AB}\) as the density operator of the initial ancilla-system state. We choose to represent \(\hat{\rho}_{AB}\) with a four-by-four matrix, \(\tau_{in}\), whose elements are given by Eq. (\ref{initial state}). Similarly, we can represent the final ancilla-system state as a matrix \(\tau_{out}\). Then, according to Eq. (\ref{tau to tau'}), the procedure of finding \(\Tilde{\chi}\) is essentially a matrix inversion:
\begin{equation}
\label{chisuper}
\tau_{out} = \tilde{\chi}\tau_{in} \longleftrightarrow
\tau_{out}\tau_{in}^{-1}=\tilde{\chi}.
\end{equation}
Now, to account for the uncertainty in the final state, \(\tau_{out}\), we define a \(4\times4\) matrix, \(\delta\tau_{out}\). Each entry of \(\delta\tau_{out}\) is the uncertainty of the corresponding entry of \(\tau_{out}\) given by its variance. Then, the Eq. (\ref{chisuper}) becomes:
\begin{equation}
\label{err}
\tau_{out} \tau^{-1} + \delta\tau_{out} \tau^{-1} = \Tilde{\chi} + \delta\Tilde{\chi}.
\end{equation}
Hence, the error matrix of the process is defined to be:
\begin{equation}
\label{err'}
    \delta\Tilde{\chi} = \delta\tau_{out} \tau_{in}^{-1}.
\end{equation}
From Eq. (\ref{err'}), we see that the error matrix of the process is equal to the final state error, \(\delta\tau_{out}\), modified by the inverse of \(\tau_{in}\). We further define the average error of the process to be:
\begin{equation}
\begin{split}
    \overline{\delta\tilde{\chi}} & = \sqrt{\frac{1}{16}\sum_{i,j=1}^4 [(\delta\tilde{\chi})_{ij}]^2} \\
      & = \sqrt{\frac{1}{16} \Tr{(\tau_{in}^{-1})(\tau_{in}^{-1})^T(\delta\tau_{out})^T(\delta\tau_{out})}}\\
     & = \frac{1}{4} \lVert (\delta\tau_{out}) ( \tau _{in}^{-1})\rVert_F\\
     & = \frac{1}{4}\lVert \delta\tilde{\chi}\rVert_F,
\end{split}
\end{equation}
where \(\lVert . \rVert_F\) is  the Frobenius norm. 
Similarly, we define the average error for \(\delta\tau_{out}\) to be:
\begin{equation}
\begin{split}
    \overline{\delta\tau_{out}} 
    & = \frac{1}{4}\lVert (\delta\tilde{\chi}) ( \tau _{in})\rVert_F\\
    & = \frac{1}{4}\lVert\delta\tau_{out}\rVert_F.
\end{split}
\end{equation}
Let \(\lVert . \rVert_{2}\) denote the spectral norm of a matrix. Using the matrix inequality theorem in ref. \cite{fang}, we can give an upper bound for the average error \(\overline{\delta\tilde{\chi}} \) and for \(\lVert\tau_{out}\rVert_F\):
\begin{equation}
\label{svb}
\begin{split}
&\overline{\delta\tilde{\chi}}=\frac{1}{4} \lVert (\delta\tau_{out}) ( \tau _{in}^{-1})\rVert_F \leq \overline{\delta\tau_{out}} \lVert \tau _{in}^{-1}\rVert_{2},\\
&\lVert\tau_{out}\rVert_F\leq\lVert\tilde{\chi}\rVert_{F}\lVert\tau_{in}\rVert_{2}.
\end{split}
\end{equation}
Combining the two inequalities in Eq. (\ref{svb}), we reach the final inequality that compares the value of \(\overline{\delta\tilde{\chi}}/\lVert\tilde{\chi}\rVert_F\) and \(\overline{\delta\tau_{out}}/\lVert\tau_{out}\rVert_F\):
\begin{equation}
\frac{\overline{\delta\tilde{\chi}}}{\lVert\tilde{\chi}\rVert_F} \leq \lVert\tau_{in}\rVert_{2}\lVert\tau_{in}^{-1}\rVert_{2} \frac{\overline{\delta\tau_{out}}}{\lVert\tau_{out}\rVert_F}.
\end{equation}
Let {\(\lambda_i\)} be the singular value of \(\tau_{in}\). The quantity \(\kappa(\tau_{in})\) defined as:
\begin{equation}
\kappa(\tau_{in})=\lVert\tau_{in}\rVert_{2}\lVert\tau_{in}^{-1}\rVert_{2} = \frac{\lambda_{max}}{\lambda_{min}},
\end{equation}
is the condition number for \(\tau_{in}\). It indicates the potential error amplification in determining \(\tilde{\chi}\) from making measurements on \(\tau_{out}\). When \(\lambda_{max} = \lambda_{min}\), there is no amplification of the relative error. Therefore, when seeking an initial state that optimizes AAPT, the states with equal singular values are preferable.

Let's first consider the case in which \(\tau_{in}\) is diagonal. Any state that takes this form belongs to a class called the X state \cite {quesada, Zhao, James}. Let's define {\(\lambda_i\)} to be the diagonals of \(\tau_{in}\). We assume the process is lossless, so \((\tau_{in})_{00} = \Tr{\hat{\rho}\hat{I}\otimes\hat{I}}=1\), and \(\lambda_1=1\). Then, we find diag(\(\tau_{in}^{-1}\)) = (\(1,\lambda_2^{-1},...,\lambda_n^{-1}\)). According to our condition number analysis, the smallest error amplification occurs when \(\lambda_2=...=\lambda_n\). Recall that we can always write \(\vert\det(\tau_{in})\vert=\lambda_1\lambda_2...\lambda_n
\).
Thus, the optimal X states are those with diagonal values, diag\((1, \vert\det(\tau_{in})\vert^{1/(n-1)}, ..., \vert\det(\tau_{in})\vert^{1/(n-1)})\). The condition number for the set of optimal states is:
\begin{equation}\label{iso}
    \kappa(\tau_{in}) = \frac{1}{\vert \det(\tau_{in})\vert^{1/n-1}}.
\end{equation}
Hence, we see that the absolute value of the Sinisterness of the state provides a numerical estimation of how well an X state is suited for performing AAPT in the best-case scenario.

The second type of states we would like to examine is the isotropic states. These states satisfy:
\begin{equation}
\Tr[\hat{\rho}(\hat{I}\otimes\sigma_i)]=\Tr[\hat{\rho}(\sigma_i\otimes \hat{I})] = 0, 
\end{equation}
for i = 1,2,3 in the two-qubit case. And so, \(\tau_{in}\) has the form:
\begin{equation}
    \tau_{in} = \begin{pmatrix}
        1 &0&0&0\\
        0&\tau_{11}&\tau_{12}&\tau_{13}\\
        0&\tau_{21}&\tau_{22}&\tau_{23}\\

        0&\tau_{31}&\tau_{32}&\tau_{33}\\
    \end{pmatrix}.
\end{equation}
These states can be mapped to the corresponding X state by performing singular value decomposition. In other words, if we choose to make measurements on an appropriate new basis, we recover the statistics of the X states. For higher dimension, the isotropic state satisfies: \(\Tr[\hat{\rho}(\hat{E}_i\otimes\hat{I})] = \Tr[\hat{\rho}(\hat{I}\otimes\hat{E}_i)] = 0\) for the choice of basis \(\hat{E}_i\). Following our argument for X states, the smallest possible condition number is again \(1/\vert \det(\tau_{in})\vert^{1/(n-1)}\), and the Sinisterness gives the best-case estimation of error amplification.

For a more general \(\tau_{in}\), which is not isotropic, the condition number does not directly relate to the Sinisterness. Rayleigh quotients could be used to estimate the maximal singular value \cite{Rayleigh}. The condition number satisfies the inequality below:
\begin{equation}
\label{kappa}
\kappa(\tau_{in}) \geq\frac{\lambda_{max}}{(\vert\det(\tau_{in})\vert/\lambda_{max})^{1/n-1}} = \frac{\lambda_{max}^{n/(n-1)}}{\vert\det(\tau_{in})\vert^{1/n-1}}.
\end{equation}
In this case, we can only conclude that the Sinisterness is inversely proportional to the condition number. In Appendix B, we show that for a bipartite system consisting of a single system qubit with an ancilla qubit, the Frobenius norm is always higher for a nearly isotropic state compared to an X state with the same determinant. Consider the solution for the minimization problem:
\begin{equation}\label{minimize}
\begin{split}
    fix: V = \lambda_1\lambda_2\lambda_3\lambda_4\\
    minimize: \sum_{i=1}^4 \lambda_i^2.
\end{split}
\end{equation}
The minimum is attained when all \(\lambda_i\) are equal. The greater Frobenius norm implies the possibility of seeking an isotropic state with a smaller difference between the largest and the smallest singular value. Hence, we expect the isotropic or corresponding X state to be a local minimum for the condition number.

Finally, we would like to comment on some other parameters used to describe the sensitivity of the initial state toward the error in \(\tau_{out}\). The commonly used definition of faithfulness in optics, given in ref. \cite{Altepeter, Mohseni2008}, is that the state is faithful if it has a maximal Schmidt number. A measure is given by \(F(\tau_{in})=\sum_l \lambda_l^2 = \Vert \tau_{in} \Vert_F\), with \(\lambda_l\) the coefficients of Schmidt decomposition of the state equal to the singular values of \(\tau_{in}\) and \(F(\tau_{in})\) is equivalent to the purity of the state \cite{Ariano}. The measure above is based on the fact that error amplification is inversely proportional to the singular values of the matrix. A larger \(F(\tau_{in})\) imply a smaller error amplification in the sense that the total singular values are larger and so their inverse should be small. We need to note that the situation is different from our minimization problem in Eq. (\ref{minimize}) since this parameter potentially compares states with different \(\vert\det(\tau_{in})\vert\). Nevertheless, \(F(\tau_{in})\) only accounts for the total amount of the error. It disregards the possibility that the singular values are unevenly distributed, which could increase the condition number. Also, it does not tell us whether the state is faithful; one still needs to perform a full singular value decomposition. Alternatively, we can write the measure of faithfulness as \(\Vert \tau^{-1}_{in} \Vert_F = \sqrt{\sum_i \Pi_{j\neq i}\lambda_j} / \vert \det(\tau_{in}) \vert = \Vert Adj(\tau_{in}) \Vert_F / \vert \det(\tau_{in}) \vert\) when \(\vert \det(\tau_{in})\vert \neq 0\), where \(Adj(\tau_{in})\) is the adjugate of \(\tau_{in}\). This shows that we can isolate the problem of determining the faithfulness of the state from performing the Schmidt decomposition. Sinisterness is straightforward to calculate; one only needs to examine the determinant of the core matrix given in ref. \cite{James}. In the two-qubit scheme, this reduces the problem of finding the set of singular values of a four-by-four matrix to the problem of finding the determinant of a three-by-three core matrix. The perspective of generalizing the quantity to a larger bipartite system means this simplification can be potentially significant. To fully characterize the error amplification of a non-isotropic state, one still needs to perform the full singular value decomposition. Sinisterness's ability to identify faithful two-qubit states from non-faithful ones demonstrated its function as an indicator of non-classical correlation \cite{James}. 

\section{Separable Mixed States, and Pure Entangled States} 
\subsection{The Separable Mixed States}
Now, we want to apply the concept of Sinisterness to separable mixed states to observe the advantage of entanglement. We begin by writing a separable mixed state in its optimal decomposition \cite{DiVincenzo}, which means writing it in the decomposition with maximal cardinality \cite{Wootters}. We write the state in the decomposition with as few distinct pure states as possible. In the case of two qubits, the maximal cardinality is four \cite{James}, i.e., any system-ancilla state can be written as:
\begin{equation}
\hat{\rho}=\sum_{n=1}^{4} P_n \hat{\rho}^{(A)}_n\otimes\hat{\rho}^{(B)}_n.
\end{equation}
Here, the \(\hat{\rho}^{A}_n\), \(\hat{\rho}^{B}_n\) are the pure state density matrices of the system and the ancilla. We do not require them to be in the same state with the same subscript n, nor do we require them to be orthogonal with different subscripts. 

Define \(Tr(\hat{\rho}^{(A)}_n \boldsymbol{\sigma})=\boldsymbol{a_n}\), \(Tr(\hat{\rho}^{(B)}_n \boldsymbol{\sigma})=\boldsymbol{b_n}\).
The calculations and simplifications are provided in the reference \cite{James} Appendix. The determinant would be:
\begin{equation}
\label{mixsep}
\det(\tau)= 36P_1P_2P_3P_4V(\boldsymbol{a_{1}}, \boldsymbol{a_2}, \boldsymbol{a_3}, \boldsymbol{a_4}) V(\boldsymbol{b_1}, \boldsymbol{b_2}, \boldsymbol{b_3}, \boldsymbol{b_4} ).
\end{equation}
In the above expression, \(V(\boldsymbol{a_{1}}, \boldsymbol{a_2}, \boldsymbol{a_3}, \boldsymbol{a_4})\) is the volume of a tetrahedron with vertices \(\boldsymbol{a_{1}}, \boldsymbol{a_2}, \boldsymbol{a_3}, \boldsymbol{a_4}\) contained on a unit sphere. We obtain a connection between the faithfulness of the state, the determinant of a matrix representing the state, and a geometric representation of the state in the Bloch sphere. By maximizing the volume of the tetrahedron inscribed in a unit sphere, we find that the value of the determinant is bound by:
\begin{equation}
\label{bound}
-\frac{1}{27} \leq det(\tau) \leq \frac{1}{27}.
\end{equation}

For the state to be faithful, we would require the determinant to be non-zero. Hence, all the \(P_n\) need to be non-zero. This means that when searching for a two-qubit separable mixed state suited for AAPT, we must choose states with maximal cardinality. We also want to point out that for a separable mixed state to saturate the bound given by Eq. (\ref{bound}), we need \(P_n=\frac{1}{4}\) and the vertices in Eq. (\ref{mixsep}) forms a regular tetrahedron, and so, the \(\Vec{a}=\sum_{n=1}^4 P_n\Vec{a_n}=0\) and \(\Vec{b}=\sum_{n=1}^4 P_n\Vec{b_n}=0\). Interestingly, this configuration corresponds to a situation in which the total state is isotropic and the component states are the most indistinguishable from one another on average. In a sense, this maximizes the quantumness arising from the basis operators' non-commutative nature. 

Using the expression of the condition number provided by Eq. (\ref{iso}), for the Werner states and the isotropic separable mixed state made of two qubits, the condition number is:
\begin{equation}
    \kappa(\tau_{in}) = 3.
\end{equation}
According to our analysis from Section 3, a small perturbation away from these states results in an increasing condition number. 

In Appendix A, we present the derivation for the Sinisterness with \(N\geq2\) system qubits, the isotropic and Werner states has \(\vert\det(\tau_{in})\vert\) scales with \(1/(4^N)^{4^N}\), hence indicating an extremely large error amplification. The condition number is given by:
\begin{equation}
    \kappa(\tau_{in}) = 4^N-1.
\end{equation}

\subsection{The Comparison between the Pure Entangled State and the Separable Mixed States}
From Section 3, we see that det(\(\tau_{in}\))=\(-\)C\(^4\), where C is the concurrence of the states which describe the amount of entanglement. For all pure states, entanglement is necessary for the state to be faithful. Assuming \(\tau_{in}\) is diagonal and the error amplification is minimized, we can assign diagonal values 1, -\(C^{4/3}\), \(C^{4/3}\), \(C^{4/3}\). If we seek the smallest error amplification, we arrive at the collection of maximally entangled states with det(\(\tau_{in}\))=\(-1\).

For separable mixed states, det(\(\tau_{in}\)) is given by Eq. (\ref{mixsep}). From the expression, we know that all faithful separable mixed states need to saturate their cardinality. Furthermore, the Bloch vectors of the four terms in the convex combination must form a tetrahedron of non-trivial volume. With the bound given by Eq. (\ref{bound}), we can estimate the smallest amount of error amplification associated with separable mixed states, diag(\(\tau_{in}\)) = ( 1, -1/3, 1/3, 1/3), which includes the separable Werner state. Alternatively, we can choose the diagonal as 1, 1/3, 1/3, and 1/3, corresponding to a mixed product state. Both states have the same condition number \(\kappa(\tau_{in})=3\). Our results suggest that an improvement in error amplification due to quantum entanglement is more prominent at a Sinisterness magnitude greater than \(\frac{1}{27}\). Lastly, the Sinisterness provides alternative evidence that an inverse-free strategy for AAPT does not exist. An inverse free state gives rise to a state matrix \(\tau_{in} = I\), which is associated with the Sinisterness value of one. However, Sinisterness takes on values ranging from -1 to \(\frac{1}{27}\), and so it automatically forbids such an inverse-free strategy.

\section{Conclusion} \label{sec:con}
We analyzed the ancilla-assisted process tomography procedure for a single qubit undergoing an unknown process. We demonstrated Sinisterness's ability to distinguish between faithful and non-faithful states. Since the success of AAPT depends on non-classical correlations, we demonstrated its function as an indicator of such non-classical correlations. We analyze the error amplification due to matrix inversion in terms of the condition number of \(\tau_{in}\). We further discussed this quantity's connection with the absolute value of Sinisterness \(\vert \det(\tau_{in})\vert\). We show that the condition number satisfy \(\kappa(\tau_{in}) \geq \lambda_{max}^{4^N/4^N-1}/\vert\det(\tau_{in})\vert^{1/(4^N-1)}\). As for two-qubit separable mixed states, which exhibit non-classical correlation but no entanglement, we represented them with tetrahedra inscribed in a unit sphere. By maximizing their volume, we find the bound on the absolute value of their Sinisterness to be 1/27. The Sinisterness takes on values ranging from [-1,0) for an entangled pure state. A comparison between entangled and separable mixed states reveals that to gain a significant advantage from entanglement, one needs the state to have a Sinisterness magnitude greater than 1/27. The above completes the discussion for characterizing the faithfulness of separable mixed states for two qubits. Finally, although the analysis of the Sinisterness of the state in this paper is based on two-qubit systems, the concept of Sinisterness can be defined in arbitrary-sized bipartite systems, since it is more generally defined as the determinant of the density operator projected in an orthogonal basis. In \(N\geq 2\) system state, the bound of Sinisterness can be found for isotropic and Werner states that consist of qubits, as in Appendix A.

\appendix
\section{Separable Mixed State with More Qubits}
To begin with, we still need to choose an orthonormal basis for describing a bipartite system containing \(N\geq2\) system qubits and N ancilla qubits. Let the basis be \{\(\hat{E_i}\otimes \hat{E_j}\)\} with i, j = [\(1,..., 2^N\)]. Where the \(\hat{E_i}\) satisfies Tr\{\(\hat{E_i}\hat{E_j}\)\}= \(\delta_{ij}\). Let M = \(4^N\) so the full state would be described by an M by M matrix in such a basis:
\begin{equation}
    \hat{\rho} = \sum_{i,j = 1}^M \tau_{ij} \hat{E_i}\otimes\hat{E_j}
\end{equation}
The same principle can be applied to this state as in the two-qubit case to decompose the matrix into L, C, and R, reducing the correlations to an (M-1) by (M-1) matrix. Define \(\Vec{a}= Tr\{\hat{\rho}\hat{\Vec{E}}\otimes \hat{\mathcal{I}}\}\) and \(\Vec{b}= Tr\{\hat{\rho} \hat{\mathcal{I}}\otimes\hat{\vec{E}}\}\), we can write \(C_{ij} = \tau_{ij}-a_ib_j\). Like before, the Sinisterness of a state is:
\begin{equation}
    det(\tau) = det (C)
\end{equation}
The cardinality of separable mixed states in higher dimensions doesn't yet have a general formula. However, some bounds have already been investigated. It has been shown that the cardinality, \(L(\hat{\rho})\), is lower bounded by the rank of the matrix representation of the state \cite{DiVincenzo}. Additionally, the decomposition has been analyzed for the isotropic and Werner states, which consist of qubits \cite{Li}. To perform AAPT, we require the matrix representation to have a full rank, meaning that a faithful separable mixed bipartite state needs to have a minimal cardinality M for a state represented by an M by M matrix. Let's investigate this case and estimate the maximum Sinisterness for the state. Let the state be:
\begin{equation}
    \hat{\rho} = \sum_{i=1}^M P_i \vert \psi_i \rangle \langle \psi_i \vert \otimes \vert \phi_i \rangle \langle \phi_i \vert
\end{equation}
Define \(\vec{a}_n = \langle \psi_n \vert \hat{\vec{E}} \vert \psi_n \rangle\), and \(\vec{b}_n = \langle \phi_n \vert \hat{\vec{E}} \vert \phi_n \rangle\). Also define \(\vec{\Bar{a}}=\sum_n P_n \vec{a}_n\) and \(\vec{\Bar{b}}=\sum_n P_n \vec{b}_n\). Following the notation for two qubits:
    \begin{equation}
    \begin{split}
    det(\tau)= \frac{1}{(M-1)!}\sum_{n_1,n_2,...,n_{M-1}=1}^M P_{n_1}...P_{n_{M-1}}(\vec{a}_{n_1}-\vec{\Bar{a}})\wedge...\wedge(\vec{a}_{n_{M-1}}-\vec{\Bar{a}})\\(\vec{b}_{n_1}-\vec{\Bar{b}})\wedge...\wedge(\vec{b}_{n_{M-1}}-\vec{\Bar{b}})
    \end{split}
\end{equation}
Similar to the two-qubit case, we can write:
    \begin{equation}
    (\vec{a}_{n_1}-\vec{\Bar{a}})\wedge...\wedge(\vec{a}_{n_{M-1}}-\vec{\Bar{a}}) = \sum_{l=1}^M P_l(\vec{a}_{n_1}-\vec{a}_{n_2})\wedge...\wedge(\vec{a}_{n_{M-1}}-\vec{a}_l)
\end{equation}
We note that the volume of a (M-1)-simplex with vertices given by \{\(\vec{a}_{n_i}, \vec{a}_l\)\} is:
    \begin{equation}
    V(\vec{a}_{n_1},...,\vec{a}_{n_{M-1}},\vec{a}_l) = \frac{1}{(M-1)!} (\vec{a}_{n_1}-\vec{a}_{n_2})\wedge...\wedge(\vec{a}_{n_{M-1}}-\vec{a}_l)
\end{equation}
Again writing this quantity with permutation indicated by \(s_{n_1,...n_{M-1},n_l}\) result in:
    \begin{equation}
    V(\vec{a}_{n_1},...,\vec{a}_{n_{M-1}},\vec{a}_l) = s_{n_1,...,n_{M-1},l}  V(\vec{a}_{1},...,\vec{a}_{M-1},\vec{a}_M)
\end{equation}
And so, the determinant is given by:
\begin{equation}
\begin{split}
   det(\tau) & = [(M-1)!]^2 \sum_{n_1,...,n_{M-1},l}^M P_{n_1}...P_{n_{M-1}}P_{l}^2 s_{n_1,...,n_{M-1},l}^2 \\ 
   & V(\vec{a}_{1},...,\vec{a}_{M-1},\vec{a}_M)V(\vec{b}_{1},...,\vec{b}_{M-1},\vec{b}_M)\\
    & = [(M-1)!]^2 P_1...P_M V(\vec{a}_{1},...,\vec{a}_{M-1},\vec{a}_M)V(\vec{b}_{1},...,\vec{b}_{M-1},\vec{b}_M)
\end{split}
\end{equation}

The existence of such a simplex has been shown in ref. \cite{Li} for the Werner states. To maximize the determinant above, choose all the \(\{P_i\}\) equal and the (M-1) simplex to be a regular simplex \cite{Fejes}. These two regular simplexes are inscribed in a unit hypersphere due to the normalization of the basis we used. For a regular n-simplex with side length d, the volume V is :
\begin{equation}
    V = \frac{d^n}{n!}\sqrt{\frac{n+1}{2^n}}
\end{equation}
The side length of a regular n-simplex inscribed in a unit hypersphere is:
\begin{equation}
    d = \sqrt{2+\frac{2}{n-1}} = \sqrt{\frac{2n}{n-1}}
\end{equation}
Combine the equations above, and one gets:
\begin{equation}
\begin{split}
    \vert det(\tau) \vert &= [(M-1)!]^2\frac{1}{M^M} \frac{M^M}{[(M-1)!]^2(M-1)^{M-1}}\\
    & = \frac{1}{(M-1)^{(M-1)}}
\end{split}
\end{equation}
This situation corresponds to a diagonal \(\tau\) with entries (1, 1/(M-1),...,1/(M-1)), the error amplification is characterized by 1/M-1. This means the error amplification for a mixed state with cardinality M in the best-case scenario, compared to the maximally entangled state, would be M-1 times larger. Suppose the system contains N qubits, then M = \(4^N\), and so the error in using mixed state roughly goes up with \(4^N\)-1.

\section{Frobenius Norm of a Single Qubit System with an Ancilla Qubit}
Let \(\tau_{in}\) be the representation of a general state \(\hat{\rho}_{in}\) in an orthogonal basis:
\begin{equation}\label{general rho}
    \hat{\rho}_{in} = \frac{1}{2}(I+\Vec{a'}\cdot\hat{\Vec{\sigma}
    })\otimes\frac{1}{2}(I+\Vec{b'}\cdot\hat{\Vec{\sigma}}) + \sum_{ij} c_{ij} \hat{\sigma_i}\otimes\hat{\sigma_j},
\end{equation}
We can always perform the decomposition:
\begin{equation}
    (U)(R)(\tau_{in})(L)(V^T)=\tau_{x},
\end{equation}
where \(\tau_x\) is a diagonal matrix, \(\tau_{x}\) = diag(1, \(s_1\), \(s_2\), \(s_3\)) satisfying \(\vert\det(\tau_x)\vert = \vert\det(\tau_{in})\vert\). The matrices U, V, R and L have determinants equal to one. Note that the decomposition above is not a singular value decomposition. Specifically, matrix L, R has the form:
\begin{equation}
    L = \begin{pmatrix}
        1 &0&0&0\\
        -a_1'&1&0&0\\
        -a_2'&0&1&0\\
        -a_3'&0&0&1
    \end{pmatrix},
    R = \begin{pmatrix}
        1&-b_1'&-b_2'&-b_3'\\
        0&1&0&0\\
        0&0&1&0\\
        0&0&0&1
    \end{pmatrix},
\end{equation}
and U, V are unitary matrices representing proper or improper rotations with the form:
\begin{equation}
    U = \begin{pmatrix}
        1 &0&0&0\\
        0&u_{11}&u_{12}&u_{13}\\
        0&u_{21}&u_{22}&u_{23}\\
        0&u_{31}&u_{32}&u_{33}
    \end{pmatrix},
    V = \begin{pmatrix}
        1&0&0&0\\
        0&v_{11}&v_{12}&v_{13}\\
        0&v_{21}&v_{22}&v_{23}\\
        0&v_{31}&v_{32}&v_{33}\\
    \end{pmatrix}.
\end{equation}
In other words, we are rewriting the decomposition of the input state in Eq. (\ref{general rho}) in another basis that resulted from rotations U and V:
\begin{equation}
\label{nb}
    \hat{\rho}_{in} = \frac{1}{2}(I+\Vec{a}\cdot\hat{\Vec{\alpha}
    })\otimes\frac{1}{2}(I+\Vec{b}\cdot\hat{\Vec{\beta}}) + \sum_{i} s_{i} \hat{\alpha_i}\otimes\hat{\beta_i}.
\end{equation}
By explicit calculation, we have:
\begin{equation}
\label{fn}
    \lVert \tau_{in} \rVert_F^2 = \lVert \tau_{x}\rVert_F^2 + \vert\vec{a}\vert^2+\vert\vec{b}\vert^2 + \vert\vec{a}\vert^2\vert\vec{b}\vert^2
    + 2\sum_{i=1}^3 a_ib_is_i.
\end{equation}

Now, let's write that:
\begin{equation}\begin{split}
    & \sum_{i=1}^3 a_ib_is_i = k\Vec{a}^T W \Vec{b} = k\vert\Vec{a}\vert \vert\Vec{b}\vert \cos{\theta},\\
    & k = \frac{\sqrt{\sum_{i=1}^3 b_i^2s_i^2}}{\vert \Vec{b} \vert},
\end{split}
\end{equation}
The matrix W is just a unitary rotation. So, Eq. (\ref{fn}) is just:
\begin{equation}
\label{fie}
\begin{split}
    \lVert \tau_{in}\rVert_F^2 & = \lVert \tau_{x}\rVert_F^2 + \vert\vec{a}\vert^2+\vert\vec{b}\vert^2 + \vert\vec{a}\vert^2\vert\vec{b}\vert^2
    + 2 k \vert\vec{a}\vert\vert\vec{b}\vert\cos(\theta)\\
    & =\lVert \tau_{x}\rVert_F^2 + \vert\vec{a}\vert^2+\vert\vec{b}\vert^2 + \vert\vec{a}\vert^2\vert\vec{b}\vert^2
    + 2 k \vert\vec{a}\vert\vert\vec{b}\vert\cos(\theta) + k^2\vert\vec{b}\vert^2\cos(\theta)^2-k^2\vert\vec{b}\vert^2\cos(\theta)^2\\
    & = \lVert \tau_{x}\rVert_F^2 + (\vert\vec{a}\vert+k\vert\Vec{b}\vert \cos{\theta})^2+\vert\vec{b}\vert^2(1+\vert\vec{a}\vert^2-k^2\cos(\theta)^2).
\end{split}
\end{equation}
As long as an X state with \(\tau_x=diag(1,s_1,s_2,s_3)\) exist, it is guaranteed that, \(1 +\vert\vec{a}\vert^2-k^2\cos(\theta)^2\geq 0\) with \(\vert \vec{a}\vert\) and \(\vert \vec{b}\vert\) small. Therefore, we find a lower bound for \(\lVert \tau_{in}\rVert_F\) for states that are nearly isotropic:
\begin{equation}
\label{tau>taux}
    \lVert \tau_{in}\rVert_F \geq \lVert \tau_{x}\rVert_F.
\end{equation}
In the above inequality, the bound is attained only when the state is exactly an X state.


\nocite{*}
\begin{thebibliography}{}
\bibitem{Chuang}
I. L. Chuang, M. A. Nielsen, J. Mod. Opt. 44, 2455–2467 (1997).

\bibitem{Poyatos}
J. F. Poyatos, J. I. Cirac, and P. Zoller, Phys. Rev. Lett. 78, 390 (1997).

\bibitem{Pryde}
J. L. O'Brien, G. J. Pryde, A. Gilchrist, D. F. V. James, N. K. Langford, T. C. Ralph, and A. G. White, Phys. Rev. Lett. 93(10), 080502 (2004).

\bibitem{ShuklaMhesh}
A. Shukla and T. S. Mahesh, Phys. Rev. A 90, 052301
(2014).

\bibitem{Mohseni2006}
M. Mohseni and D. A. Lidar, Phys. Rev. Lett. 97, 170501
(2006)

\bibitem{Mohseni2007}
M. Mohseni and D. A. Lidar, Phys. Rev. A 75, 062331 (2007)

\bibitem{Mohseni2008}
M. Mohseni, A. T. Rezakhani, and D. A. Lidar, Phys. Rev. A, 77, 032322 (2008).

\bibitem{Altepeter}
J. B. Altepeter, D. Branning, E. Jeffrey, T. C. Wei, P. G. Kwiat, R. T. Thew, J. L. O’Brien, M. A. Nielsen, and A. G. White, Phys. Rev. Lett. 90, 193601 (2003).

\bibitem{Modi}
K. Modi, T. Patrek, W. Son, V. Vedral and M. Williamson, “Unified View of Quantum and Classical Correlations,” Phys. Rev. Lett. 104, 080501 (2010) (4pp)

\bibitem{Ariano}
G. M. D’Ariano and P. Lo Presti, Phys. Rev. Let. 86, 4195 (2001)

\bibitem{TP}
Thomas, P., Ruscio, L., Morin, O. et al., Efficient generation of entangled multiphoton graph states from a single atom. Nature 608, 677–681 (2022). 

\bibitem{ZL}
Zhou, L., Liu, ZK., Xu, ZX. et al. Economical multi-photon polarization entanglement purification with Bell state. Quantum Inf Process 20, 257 (2021). 

\bibitem{VS}
Varo, S., Juska, G., Pelucchi, E. An intuitive protocol for polarization-entanglement restoral of quantum dot photon sources with non-vanishing fine-structure splitting. Sci Rep 12, 4723 (2022). 

\bibitem{SHL}
Seok Hyung Lie and Hyunseok Jeong, Phys. Rev. Lett. 130, 020802 (2023)

\bibitem{LZD}
G.-D. Lu, Z. Zhang, Y. Dai, Y.-L. Dong, and C.-J. Zhang,
Ann. Phys. (Amsterdam) 534, 2100550 (2022).

\bibitem{Wood}
C. J. Wood, J.D. Biamonte, and D.G. Cory, Quantum Inf. Comput. 15, 759 (2015).

\bibitem{choi}
M.-D. Choi, Linear Algebra Appl. 10, 285 (1975). 

\bibitem{jam}
A. Jamiołkowski, Rep. Math. Phys. 3, 275 (1972).

\bibitem{quesada}
Quesada, N., Al-Qasimi, A., James, D. F. V. (2012). Quantum properties and dynamics of X states. Journal of Modern Optics, 59(15), 1322–1329. https://doi.org/10.1080/09500340.2012.713130

\bibitem{Zhao}
Zhao, MJ., Ma, T., Wang, Z.et al., Coherence concurrence for X states., Quantum Inf Process, 19, 104 (2020). https://doi.org/10.1007/s11128-020-2601-2

\bibitem{James}
D. F. V. James, Journal of the Optical Society of America A, Vol. 39, Issue 12, pp. C86-C97 (2022).

\bibitem{DiVincenzo}
D.P. DiVincenzo, B.M. Terhal, and A.V. Thapliyal, “Optimal decomposition of barely separable states,” J. Mod. Opt. 47, 377-385 (2000).

\bibitem{Wootters}
W. K. Wootters, “Entanglement of Formation of an Arbitrary State of Two Qubits,” Phys. Rev. Lett. 80(10),
2245-2248 (1998).

\bibitem{fang}
Y. Fang, K. A. Loparo and Xiangbo Feng, "Inequalities for the trace of matrix product," in IEEE Transactions on Automatic Control, vol. 39, no. 12, pp. 2489-2490, Dec. 1994, doi: 10.1109/9.362841.

\bibitem{Fejes}
Fejes T´ot, L., Regular Figures, New York: Macmillan/Pergamon, 1964.

\bibitem{Li}
Li, JL., Qiao, CF. A Necessary and Sufficient Criterion for the Separability of Quantum State. Sci Rep 8, 1442 (2018). https://doi.org/10.1038/s41598-018-19709-z

\bibitem[]{Belsley}\
Belsley, D. A., Kuh, E. and Welsch, R. E. (2004). Regression diagnostics: identifying influential data and sources of collinearity. Wiley.

\bibitem[]{Rayleigh}
Horn RA, Johnson CR. Hermitian and symmetric matrices. In: Matrix Analysis. Cambridge University Press; 1985:167-256.
\end{thebibliography}
\end{document}